Towards reliable head and neck cancers locoregional recurrence prediction using delta-radiomics and learning with rejection option

Kai Wang, Michael Dohopolski
*Department of Radiation Oncology, UT Southwestern Medical Center, Dallas, TX 75390, USA*

Qiongwen Zhang
*Department of Head and Neck Oncology, State Key Laboratory of Biotherapy and Cancer Center, West China Hospital, Sichuan University, Chengdu 610041, China*
*Department of Radiation Oncology, UT Southwestern Medical Center, Dallas, TX 75390, USA*

David Sher and Jing Wang[a]
*Department of Radiation Oncology, UT Southwestern Medical Center, Dallas, TX 75390, USA*

[a] Author to whom correspondence should be addressed. Email: Jing.Wang@UTSouthwestern.edu

**Purpose:** A reliable locoregional recurrence (LRR) prediction model is important for the personalized management of head and neck cancers (HNC) patients who received radiotherapy. This work aims to develop a delta-radiomics feature-based multi-classifier, multi-objective, and multi-modality (Delta-mCOM) model for post-treatment HNC LRR prediction. Furthermore, we aim to adopt a learning with rejection option (LRO) strategy to boost the reliability of Delta-mCOM model by rejecting prediction for samples with high prediction uncertainties.

**Methods:** In this retrospective study, we collected PET/CT image and clinical data from 224 HNC patients who received radiotherapy (RT) at our institution. We calculated the differences between radiomics features extracted from PET/CT images acquired before and after radiotherapy and used them in conjunction with pre-treatment radiomics features as the input features. Using clinical parameters, PET radiomics features, and CT radiomics features, we built and optimized three separate single-modality models. We used multiple classifiers for model construction and employed sensitivity and specificity simultaneously as the training objectives for each of them. Then, for testing samples, we fused the output probabilities from all these single-modality models to obtain the final output probabilities of the Delta-mCOM model. In the LRO strategy, we estimated the epistemic and aleatoric uncertainties when predicting with a trained Delta-mCOM model and identified patients associated with prediction of higher reliability (low uncertainty estimates). The epistemic and aleatoric uncertainties were estimated using an AutoEncoder-style anomaly detection model and test-time augmentation (TTA) with predictions made from the Delta-mCOM model, respectively. Predictions with higher epistemic uncertainty or higher aleatoric uncertainty than given thresholds were deemed unreliable, and they were rejected before providing a final prediction. In this study, different thresholds corresponding to different low-reliability prediction rejection ratios were applied. Their values are based on the estimated epistemic and aleatoric uncertainties distribution of the validation data.

**Results:** The Delta-mCOM model performed significantly better than the single-modality models, whether trained with pre-, post-treatment radiomics features or concatenated BaseLine and Delta-Radiomics Features (BL-DRFs). It was numerically superior to the PET and CT fused BL-DRF model (nonstatistically significant). Using the LRO strategy for the Delta-mCOM model, most of the evaluation metrics improved as the rejection ratio increased from 0% to around 25%. Utilizing both epistemic and aleatoric uncertainty for rejection yielded nonstatistically significant improved metrics compared to each alone at approximately a 25% rejection ratio. Metrics were significantly better than the no-rejection method when the reject ratio was higher than 50%.

**Conclusions:** The inclusion of the delta-radiomics feature improved the accuracy of HNC LRR prediction, and the proposed Delta-mCOM model can give more reliable predictions by rejecting predictions for samples of high uncertainty using the LRO strategy.

## 1. INTRODUCTION

Locoregional recurrence (LRR) is the predominant pattern of relapse following radiation therapy for patients with head and neck cancer (HNC).[1,2] Accurately identifying HNC patients at high risk of LRR after radiotherapy is of great importance to guide physicians to design personalized patient management plans, which have the potential to help improve the treatment outcome and quality of life for HNC patients.[1-4]

FDG-PET/CT plays a crucial role in identifying residual or recurrent disease following definitive radiotherapy of HNC patients.[3-7] For example, a decrease of 15-25% in the maximum standard uptake value (SUVMax) between the baseline and post-treatment scans suggests an appropriate treatment response according to the European Organization for Research and Treatment of Cancer (EORTC).[5] Additionally, recent studies have reported that no matter with or without clinical suspicion, through visual inspection and semiquantitative analyses of the increased $^{18}$F-FDG uptake in the primary tumors, experienced physicians can predict recurrence with high sensitivity and negative predictive value.[3,6,7]

Radiomics-based methods, which extract handcrafted quantitative features from radiological images and use machine learning (ML) methods for feature analysis and constructing predictive models, have shown promising performance in HNC-related tasks.[8-12] Parmar *et al.* extracted radiomics features from pre-treatment CT images and built a Cox logistic regression model for predicting local tumor control (LC) after chemoradiotherapy of HNC.[13] Their results demonstrated that some radiomics features were significantly associated with LC—the best concordance index (CI) was 0.78. Vallieres *et al.* used a radiomics-based method to develop models that predict LRR and distant metastases (DM) before HNC treatment.[10] They extracted radiomics features from FDG-PET/CT scans and used a random forest method to construct a multivariable model. Their best results were achieved when the model used features from multiple modalities. They obtained an area under the receiver operating characteristic curve (AUC) of 0.86 for DM and 0.69 for LRR. Using the same dataset as Vallieres *et al.,* we previously built a multi-classifier, multi-objective and multi-modality model (mCOM) for HNC LRR prediction.[8] In the mCOM model, multiple classifiers were used to create the model; sensitivity and specificity were considered simultaneously as the objectives to guide the model construction. Both clinical features and radiomics features extracted from various modalities were used as model inputs. The optimal mCOM model achieved an AUC value of 0.77.

Radiomics features extracted from radiological images collected at separate times with the same imaging protocol for the same patient can measure the therapy-induced changes in the tumor area. The changes in radiomics features, termed delta-radiomics features, have been associated with treatment response or outcome for serval types of cancers.[14-19] Lin *et al.* collected CT radiomics features extracted before and after neoadjuvant chemotherapy for high-grade osteosarcoma patients.[14] After calculating the delta-radiomics features, they built a delta-radiomics nomogram incorporating the radiomics signature and clinical factors for individualized pathologic response evaluation after chemotherapy. This nomogram showed good calibration and great discrimination capacity with AUC 0.843. Alahmari *et al.* investigated the predictivity of delta-radiomics features and conventional (baseline) features extracted from a serial of low-dose CT scans for predicting pulmonary nodule malignancy.[15] Their best result (AUC=0.82) was achieved when they combined delta-radiomics features with conventional features. For HNC, Morgan *et al.* extracted delta-radiomics features from cone beam CT (CBCT) acquired at different fractions of radiotherapy. They used them together with baseline CT radiomics features for primary tumor local failure prediction.[16] In their results, they showed that adding the delta-radiomics features to the baseline CT-radiomics features improved the AUC value from 0.69 to 0.77. For local advanced rectal cancer (LARC), Jeon *et al.* developed and evaluated the predictive ability of 2D and 3D delta-radiomics-based signatures for predicting treatment outcome of patients received preoperative chemoradiotherapy and surgery, the constructed delta-radiomics-based signatures were independent prognostic factors and successfully predicted the outcomes.[17] Shayesteh *et al.* found T2-weighted MRI delta-radiomics-based models significantly outperformed models built with pre- or post-treatment radiomic features for predicting the treatment response of LARC patients underwent neoadjuvant chemoradiation therapy (AUC=0.96).[18] Chen et al. proposed a CT radiomics-based model to predict the treatment response of immunotherapy for metastatic melanoma patients.[19] They found better prediction performance can be obtained when baseline radiomic features were combined with net change of radiomic features, i.e. delta-radiomics, for model construction. In this study, we hypothesize that quantizing the spatial intensity variation of a tumor using PET/CT radiomics features and tracking the therapy-induced change through delta-radiomics features will improve our ability to predict LRR following radiotherapy of patients with HNC.

In medical physics-related applications such as treatment outcome prediction, we often have a limited cohort of patients when constructing a ML model.[20-22] This limits the choice of ML methods. For example, although successful across many real-world applications, deep learning-based methods are easily overfitted to the training data when datasets are limited. On the other hand, even if a model is appropriately trained, it only observes a region surrounding the training samples in feature space. This region might not be an adequate representation of the ideal distribution of all the patients. Because of these issues, with a given smaller dataset, despite appropriate performance on the training and validation dataset, the model may not provide reliable predictions to all the testing samples, especially for those whose characteristics vary significantly from the training dataset distribution, as illustrated in Figure 1.

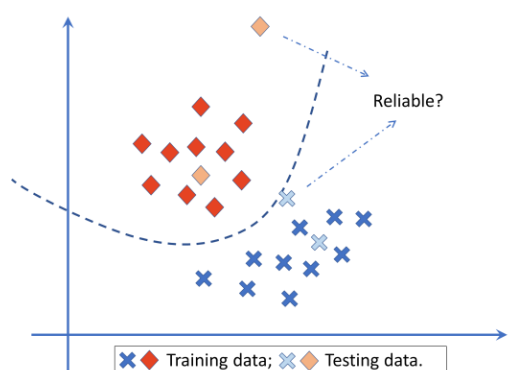

**Figure 1.** Illustration of unreliable predictions when a model is trained with limited data samples (modified from Ref. 22). When the query sample lies close to the decision boundary, the prediction is unreliable because of low prediction confidence. Additionally, the prediction can be unreliable despite a high class probability value when the query sample is far from the training data distribution.

Currently, for treatment outcome prediction, most studies have been focused on improving the overall model performance through upgrading the feature selection method, exploring different features, introducing novel classifiers, or utilizing new imaging or detection modalities.[8,12,23-27] However, fewer studies center on prediction reliabilities for individual samples. As a clinical diagnosis/prediction application, an incorrect LRR prediction for HNC patient can be costly. Therefore, the accuracy or reliability of prediction for each patient should be viewed as a more crucial requirement than giving predictions for all query patients.[22,28] In this work, to further improve the reliability of the constructed model for HNC LRR prediction, we introduced a novel flexible outcome prediction strategy that implements uncertainty estimates for each testing sample. For a given model trained on a fixed dataset, when it is used to predict the outcome of a new query patient, we calculate its prediction epistemic and aleatoric uncertainties first. Then, the model can reject to provide the prediction result based on the value of these uncertainties. For those patients for whom the model is not confident in providing a prediction then further physician scrutiny is necessary.

## 2.   MATERIALS AND METHODS

### 2.1 Patient data

This study evaluated data from 432 patients with HNC who received definitive, conventionally fractionated radiation therapy between September 2005 to November 2015 at the University of Texas Southwestern Medical Center (UTSW). The final cohort included 224 patients after excluding patients who did not have appropriate pre- and post-treatment FDG-PET/CT imaging, had a prior history of radiation therapy, or had missing relevant clinical information. Fifty-seven patients experienced LRR, and the median time from treatment completion to LRR was 258 days (interquartile range [IQR]: 125-448 days). The median time from pre-treatment FDG-PET/CT imaging to the first day of radiation therapy was 44 days (IQR: 19-63 days), the median time from treatment completion to post-treatment scan was 122 days (IQR: 87-148 days). The median follow-up duration for this study is 60 months (IQR: 13-117 months). On pre-treatment images, contours for the gross tumor volumes (GTVs) were drawn on the planning CT and were propagated and registered to the PET/CT. On post-treatment PET/CT scan, we defined suspicious tumor volumes (STVs) as areas suspicious for LRR after radiation therapy on PET/CT images, compromising GTV and its surrounding areas with SUV≥2.5. STVs were manually contoured on CT by a radiation oncologist and nuclear medicine radiologist. These patients' clinical information, including gender, HPV status, primary tumor site, T-stage, N-stage, and treatment method, were collected to help build the outcome prediction model. Patient demographics were summarized in Supplementary Table S1. To reduce the model performance variability caused by subset partition, we adopted five-fold cross-validation for training, validation, and testing—three folds for training, one for validation, and one for testing in our experiment. The outcome ratio was preserved in each fold.

## 2.2 Radiomics and delta-radiomics feature extraction

For each imaging modality, a total of 257 radiomics features were extracted using an open-source radiomics toolbox which satisfies the methodology and definitions of the Image Biomarker Standardization Initiative.[10,29] The extracted radiomics features included nine intensity features (minimum, maximum, mean, standard deviation, sum, median, skewness, kurtosis, and variance), eight geometry features (GTV volume, major diameter of GTV, minor diameter of GTV, eccentricity, elongation, orientation, bounding box volume, and perimeter of the GTV on the slice which has the biggest tumor area), and 240 texture features (nine features from the Gray-Level Co-occurrence Matrix, thirteen features from the Gray-Level Run-Length Matrix, thirteen features from the Gray-Level Size Zone Matrix and five features from the Neighborhood Gray-Tone Difference Matrix collected under six sets of different feature extraction parameters). For the details regarding the extracted features, please refer to the supplementary material. The delta-radiomics features obtained separately from the PET and CT portions of the PET/CT scan were calculated by direct subtraction of pre- and post-treatment radiomics features.[25-27] As presented in our work on CBCT based HNC treatment outcome prediction, delta-radiomics features alone do not provide superior predictive capabilities when used independently from the baseline radiomics feature (radiomics feature extracted from the previous CBCT or pre-treatment CT).[16] In this work, we used pre-treatment radiomics features and delta-radiomics features as the radiomics feature input, a vector of 514 features for each imaging modality. We termed the concatenated BaseLine and Delta-Radiomics Features as BL-DRFs in the remainder of this work.

### 2.3 Delta-radiomics based multi-classifier multi-objective and multi-modality model

In our previous work, we proposed a multi-classifier multi-objective and multi-modality (mCOM) model for pre-treatment HNC LRR prediction. We applied multiple classifiers, comprising support vector machine (SVM), discriminant analysis (DA), and logistic regression (LR) to build the model. For model optimization, we considered sensitivity and specificity simultaneously as the objectives for model training. To fully utilize the collected data and improve model robustness, we used features from multiple modalities as model inputs—i.e., clinical parameters and radiomics feature extracted from CT and PET images.[8] In this study, to construct the delta-radiomics based post-treatment HNC LRR prediction model, we modified mCOM to Delta-mCOM by replacing the radiomics features with BL-DRFs. The workflow for the Delta-mCOM model is shown in Figure 2a.

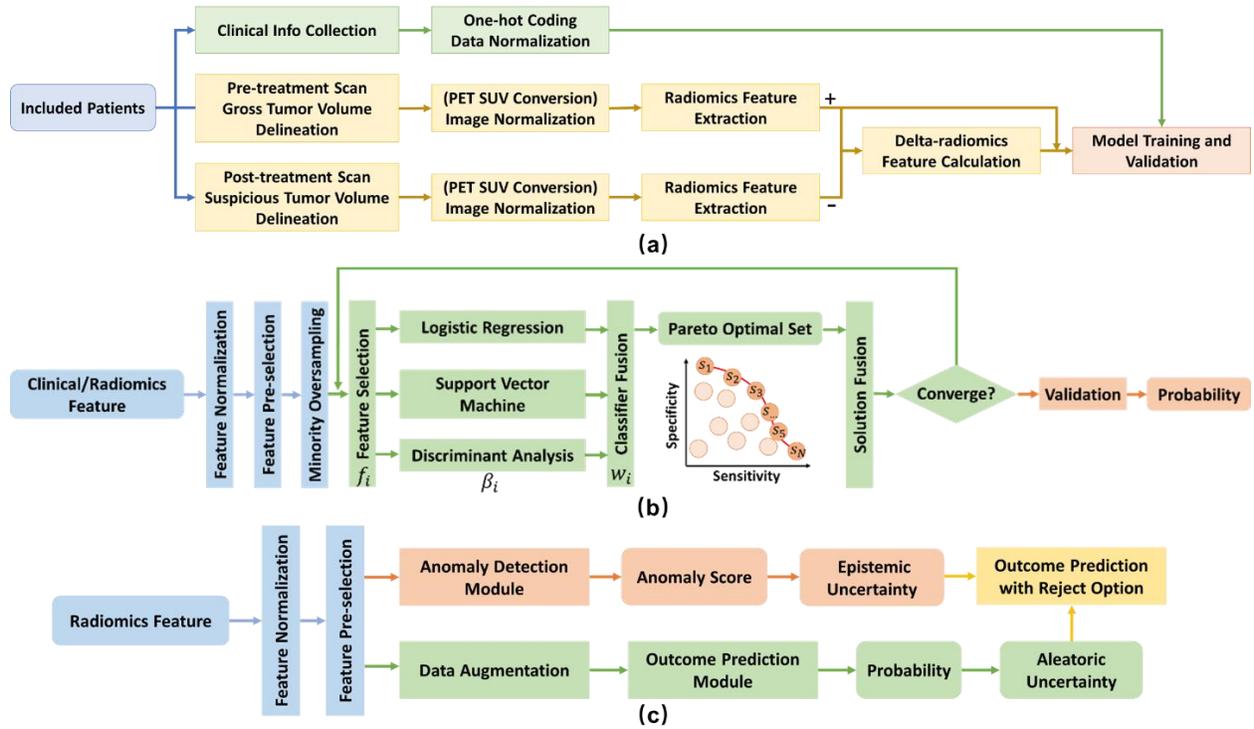

**Figure 2.** Workflow of the proposed reliable delta-radiomics feature-based multi-classifier, multi-objective, and multi-modality (Delta-mCOM) model for post-treatment head and neck cancers (HNC) locoregional recurrence (LRR) prediction. (a) Workflow of Delta-mCOM model construction. Three separated single-modality models (clinical model, PET-radiomics based model, and CT-radiomics based model) are trained individually and their out probabilities are fused together to generate the final prediction of Delta-mCOM model. (b) Workflow of the training and validation processes of a single-modality multi-objective multi-classifier prediction model. A trained model contains a set of different solutions $\{s_1, s_2, ..., s_N\}$ (Pareto-optimal solution set), each solution comprises of three trainable parameters: feature selection vector $f_i$, classifier parameter $\beta_i$, classifier weighting factor $w_i$, where $i = 1, 2, ..., I$ is the index of a solution, $I$ is the maximum size of the solution set, $N$ is the size of the Pareto-optimal solution set, and $N < I$. (c) Workflow of HNC LRR prediction using Delta-mCOM model with rejection option.

The training and validation processes of the Delta-mCOM model were the same as the mCOM model.[8] In the training stage, we optimized three single-modality models separately and fused the output probabilities from each model to obtain the final prediction in the validation/testing stages. The workflow for the training and validation/testing processes for a single-modality multi-classifier multi-objective model is shown in Figure 2b. We performed feature normalization, feature pre-selection, and minority class oversampling for training data before model training for a single-modality feature dataset. We used min-max normalization strategy for feature normalization, and the transformation parameter is based on the feature values of training cohort for each fold of cross validation. We used the minimal-redundancy-maximal-relevance criterion (mRMR) method as the feature pre-selection method to reduce the impact of feature redundancy on model training. Then we applied the synthetic minority oversampling technique (SMOTE) to oversample the LRR cohort to produce class-balanced training data.[30,31] During the training process, $I$ different solutions were randomly initialized, each of them is a set of trainable parameters: feature selection vector $f_i$, classifier parameter $\beta_i$, and classifier weighting factor $w_i$, where $i = 1, 2, ..., I$. A multi-task multi-objective immune algorithm (mTO) was used to optimize sensitivity and specificity simultaneously through iterative feature selection, classifier parameter training, evidential reasoning (ER)-based fusion of output probabilities of classifiers, Pareto-optimal solution set updating, and weighted fusion of output probabilities of Pareto-optimal solution set. For the details of mTO, ER, and weighted fusion, please refer to our previous work.[8] After $T$ generations of model updating, the single-modality model training is considered converged. For a validation or testing sample, the output probability from each single-modality model is the fused probability of its Pareto-optimal solution set output probabilities.

After the three single-modality models were trained, the Delta-mCOM model can report the prediction result of a query patient through weighted fusion of the output probabilities from clinical, PET-radiomics, and CT-radiomics based models. The weights for output probabilities for the different modality models were calculated based on the model performance on validation data.[8]

## 2.4 Learning with rejection option

While comparable to those in the literature, our dataset is undersized when accounting for the total number of radiomics features extracted. Thus, the Delta-mCOM model would perform poorly when given data sampled from regions not represented in the training dataset. Alternatively, the model might be sensitive to minor changes in the input features and can therefore be unreliable. To improve the model's reliability without collecting more data or modifying the model structure, we introduced a "learning with rejection option (LRO)" strategy, which only gives prediction results to those with low uncertainty.[22,28,32] The overall workflow for using the LRO strategy on Delta-mCOM model is shown in Figure 2c. First, we estimated epistemic and aleatoric uncertainties for a prediction to quantify the reliability of each prediction of the Delta-mCOM model. We then selected a patient group in which the model has more confidence (i.e., low epistemic or aleatoric uncertainty) and only submitted final predictions for such samples. In other words, the model would not predict whether a patient would experience a LRR if the uncertainty values were large.

### 2.4.1 Epistemic uncertainty estimation

Epistemic uncertainty of prediction refers to the uncertainty caused by insufficient knowledge regarding a corresponding application when training the prediction model, and it can be reduced by additional unique training data.[32-35] To quantify epistemic uncertainty, Jackman *et al.* proposed a method that approximated a range in the feature domain that encapsulated the bounds of the training data.[34] In the field specific to deep learning, Kendall *et al.* proposed a Bayesian deep learning framework for uncertainty estimation via dropout variation inference.[35] Anomaly detection model, also known as the outlier detection model, is widely used to identify rare items, observations, or events that differ from the general distribution.[36,37] In an unsupervised anomaly detection model, it models the data distribution using training dataset, and a testing sample that lies away from training data distribution will be associated with a higher anomaly score. Based on this, we intuitively estimated the epistemic uncertainty through measuring the anomaly scores from unsupervised anomaly detection models which were trained with the same datasets used in Delta-mCOM model training.

As radiomics features played a dominant role in predicting HNC LRR in our previous studies,[8,16] we only measured the uncertainty when PET and CT BL-DRFs were used in the Delta-mCOM model. Because PET and CT images are inherently different, and the difference in the radiomics features of positive and negative samples is the fundamental hypothesis of radiomics-based studies, we built four unsupervised anomaly detection models for modelling the training data distribution, e.g., CT-positive, CT-negative, PET-positive, and PET-negative anomaly detection models. We used the same fully connected Autoencoder (AE) structure to construct the four unsupervised anomaly detection models with four hidden layers (hidden neurons 32, 16, 16, and 32). The normalized reconstruction error was used as the anomaly score in the AE-based anomaly detection networks, 0 stands for inliers and 1 for outliers. The anomaly score here was used as the estimate epistemic uncertainty, and a high anomaly score equated to high epistemic uncertainty for the prediction.

The anomaly detection models were trained with the same BL-DRF datasets used in Delta-mCOM model training (with feature normalization and pre-selection). For a testing sample predicted to have a high probability of LRR by the Delta-mCOM model (0.5 was used as the probability threshold), this sample's epistemic uncertainty was calculated using the average value of the anomaly scores from the CT-positive and PET-positive models. And, for a testing sample predicted to be LRR negative, the epistemic uncertainty was calculated using the average value of the anomaly scores from the CT-negative and PET-negative models. The implementation of the anomaly detection model was based on an open-source python package named Python toolkit for detecting outlying objects (PyOD).[36]

### 2.4.2 Aleatoric uncertainty estimation

Aleatoric uncertainty describes the noise inherent in data and data collection, which cannot be addressed by adding more data. For deep learning-based applications, Ayhan and Berens proposed a method to estimate aleatoric uncertainty through TTA.[38] Inspired by their methods, we added zero-mean Gaussian noise with a standard deviation of 0.01 to normalized PET/CT radiomics features to obtain augmented BL-DRFs (100 alterations per PET/CT combination). After predictions were made with the augmented features, we measured uncertainty using the entropy of mean class probabilities obtained from the augmented data.[39] Mean class probability of 0.5 will have the entropy value of 1, which is the highest aleatoric uncertainty score.

### 2.4.3 Learning with rejection option For Delta-mCOM model

As radiomics features played a dominant role in predicting HNC LRR in our previous studies,[8,16] we only measured the uncertainty when PET and CT BL-DRFs were used in the Delta-mCOM model. We took consideration of these two uncertainties together for testing sample rejection. We rejected the samples with high epistemic uncertainty first, who are in theory the least similar to our training data distribution. Then we rejected samples with high aleatoric uncertainty as they may distribute close to the decision boundary or sensitive to noise when collecting data. In this study, we used different percentile values of uncertainty scores corresponding to different rejection ratios obtained from the validation dataset as the thresholds for rejection.

For example, if we were to reject predictions with an epistemic uncertainty above the 70th percentile, we would obtain that epistemic uncertainty value from the validation dataset and apply that same cut-off to the testing dataset. Around 30% of testing samples would then be expected to be rejected for giving the prediction results. And if we used the 70th percentile values for both epistemic and aleatoric uncertainties as the thresholds for the combined rejection, testing samples with epistemic or aleatory uncertainty values above their respective thresholds were rejected. In this case, the rejection ratio is expected to be higher than 30% since these two uncertainties quantify distinct aspects of prediction reliability.[33,35]

## 3. RESULTS

### 3.1 Experimental setup

In the training stage of the Delta-mCOM model, we used the same hyper-parameters as our previous work.[8] The population number $I$ was set to 100, the maximal generation number $T$ was set to 50, and the number of pre-selected features for imaging-based modalities is 50.

We designed three groups of experiments. (1) To validate the benefits of adopting multiple modalities for post-treatment HNC LRR prediction, we compared the performance of the single modality models to the multi-modality fused models. (2) To show the performance improvement by introducing delta-radiomics to HNC LRR prediction, we compared models trained with BL-DRFs to models built with separate pre-treatment and post-treatment radiomics features. (3) To verify the effectiveness of the LRO strategy, we compared the model performance without a rejection metric, epistemic uncertainty only rejection, aleatoric uncertainty only rejection, and combined rejection at different rejection ratios.

We used sensitivity, specificity, accuracy, the receiver operating characteristic curve (ROC), and AUC to evaluate the performance of the different models and different testing sample rejection methods. A probability value of 0.5 was used as the threshold to determine the prediction label. We also investigated the difference in performance between the different models and rejection methods using Delong's test with a significance level of 0.05 in R version 4.0.3.[40,41] To reduce the variability caused by subset partition, we adopted five-fold cross-validation for model training, validation, and testing. Additionally, all experiments were performed five times. The mean and standard deviation were then calculated for each evaluation criterion. In the remainder of the paper, all prediction performance reported were from the testing results.

### 3.2 Performance evaluation of multiple-modality models versus single-modality models

We trained seven single-modality models for HNC LRR prediction: clinical features model, PET pre-treatment radiomics model, PET post-treatment radiomics model, PET BL-DRF model, and CT pre-treatment radiomics model, CT post-treatment radiomics model, and CT BL-DRF model. We fused the CT and PET radiomics models to three radiomics-only models, which were CT+PET pre-treatment radiomics model, CT+PET post-treatment radiomics model, and CT+PET BL-DRF model. Correspondingly, we fused the clinical model to radiomics models to obtain the CT+PET+Clinic pre-treatment model (mCOM), CT+PET+Clinic post-treatment model, and CT+PET+Clinic BL-DRF model (Delta-mCOM).

The prediction performances of the single-modality models and the multi-modality model are shown in Table 1 and Figure 3. Whether pre-treatment or post-treatment radiomics feature were used, the fusion model of all the three modalities achieved the best performance (AUC=0.75 and 0.73 for pre- and post-treatment models, respectively). The AUC values were similar to those from the fused CT and PET models (0.74 and 0.73 for pre- and post-treatment, respectively).

### 3.3 Performance evaluation of BL-DRF model versus pre- and post-treatment radiomics models

The prediction performances of the Delta-mCOM models and models built with pre- or post-treatment features are shown in Table 1, Figures 3 and 5. All the ROCs are compared with the Delta-mCOM model using the Delong test. For single-modality models trained with PET or CT radiomics feature, BL-DRFs improved the model performance. The AUC value increased from 0.69 to 0.74 for CT-radiomics model, and from 0.72 to 0.75 for PET-radiomics model. When using BL-DRFs as the model input for the multi-modality models, the fusion of all the three modalities (Delta-mCOM) achieved the best performance on all the evaluation metrics (Sensitivity=0.75, Specificity=0.72, Accuracy=0.73, AUC=0.80). The ROC was significantly better than CT or PET BL-DRF models (AUC=0.75 and 0.74, P=0.01 and 0.01, respectively) and marginally higher than the fused CT and PET BL-DRF model (AUC=0.78, P=0.79).

**Table 1.** Results of head and neck cancers locoregional recurrence prediction with different features. P-values here are obtained by comparing the ROCs of different models to that of the proposed delta-radiomics feature-based multi-classifier, multi-objective, and multi-modality (Delta-mCOM) model (last row). Note that concatenation of baseline and delta-radiomics features (BL-DRFs) were used as the input for delta-radiomics-based models.

| Modality | Feature | Sensitivity | Specificity | Accuracy | AUC | P-value |
|----------|---------|-------------|-------------|----------|-----|---------|
| Clinic | Pre | 0.64±0.05 | 0.62±0.03 | 0.62±0.02 | 0.67±0.03 | <0.01 |
| CT | Pre | 0.69±0.03 | 0.62±0.02 | 0.64±0.01 | 0.69±0.02 | <0.01 |
| | Post | 0.62±0.03 | 0.62±0.04 | 0.62±0.03 | 0.67±0.01 | <0.01 |
| | Delta | 0.73±0.04 | 0.70±0.04 | 0.70±0.03 | 0.74±0.03 | 0.02 |
| PET | Pre | 0.65±0.04 | 0.66±0.02 | 0.66±0.02 | 0.72±0.02 | <0.01 |
| | Post | 0.65±0.03 | 0.65±0.01 | 0.65±0.01 | 0.72±0.02 | <0.01 |
| | Delta | 0.72±0.05 | 0.71±0.03 | 0.71±0.02 | 0.75±0.02 | <0.01 |
| CT+PET | Pre | 0.67±0.07 | 0.70±0.05 | 0.69±0.03 | 0.74±0.02 | <0.01 |
| | Post | 0.67±0.01 | 0.68±0.03 | 0.68±0.02 | 0.73±0.02 | 0.01 |
| | Delta | 0.71±0.03 | 0.71±0.02 | 0.71±0.02 | 0.78±0.01 | 0.79 |
| CT+PET+Clinic | Pre | 0.71±0.06 | 0.67±0.02 | 0.69±0.02 | 0.75±0.03 | 0.01 |
| | Post | 0.69±0.04 | 0.68±0.03 | 0.68±0.02 | 0.73±0.02 | <0.01 |
| | Delta | 0.75±0.04 | 0.72±0.01 | 0.73±0.01 | 0.80±0.01 | ------ |

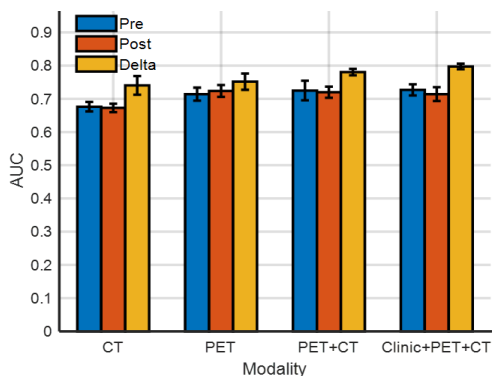

**Figure 3.** AUCs of head and neck cancers locoregional recurrence prediction with radiomics features collected from different modalities at different time points and the proposed delta-radiomics feature-based multi-classifier, multi-objective, and multi-modality (Delta-mCOM) model. Note that concatenation of baseline and delta-radiomics features (BL-DRFs) were used as the input for delta-radiomics-based models.

## 3.4 Performance evaluation of different rejection methods

The prediction performances of the Delta-mCOM models with different rejection options and rejection ratios are shown in Table 2 and Figures 4-5. The rejection ratios in Table 2 and Figures 4-5 were the actual rejection ratios calculated based on testing data. With the epistemic uncertainties (anomaly scores) and aleatoric uncertainties estimated for both the validation and testing data, we sorted the uncertainties of validation data from low to high. We used the 100th to 70th (5 as the interval) percentiles of epistemic and aleatoric uncertainties of validation data as the thresholds for testing sample rejection, corresponding to 0% to 30% rejection ratio. We applied single-uncertainty rejection and combined-uncertainty rejection with these thresholds and compared the ROCs of the different methods to that of the combined uncertainty strategy at around 25% rejection ratio and approximately 50% rejection ratio with unpaired Delong's test. For all the three rejection methods, the model performances trended towards improvement as the selected testing sample ratio decreased (i.e., rejected ratio increased). For example, at a rejection ratio of around 25%, all the rejection methods had similar performance, while the combined method (AUC=0.82) was marginally higher than the single-uncertainty rejection methods (AUCs=0.81). At a rejection ratio of around 50%, the AUC value on the selected testing data was boosted to 0.85, which was significantly higher than the performance seen on all the testing samples using the Delta-mCOM model (no-rejection method, AUC=0.80) and the performance of the Delta-mCOM model on the rejected samples (AUC=0.63). We also investigated the impact of rejection ratio to the ratio of positive samples in the selected testing data (Supplementary Figure S1). With the combined rejection method, although the LRR positive sample was the minority in our dataset, which might compromise the prediction reliability, the ratio of LRR positive samples in the selected testing group maintained a stable range (25% at 0% rejection ratio and 22% at around 50% rejection).

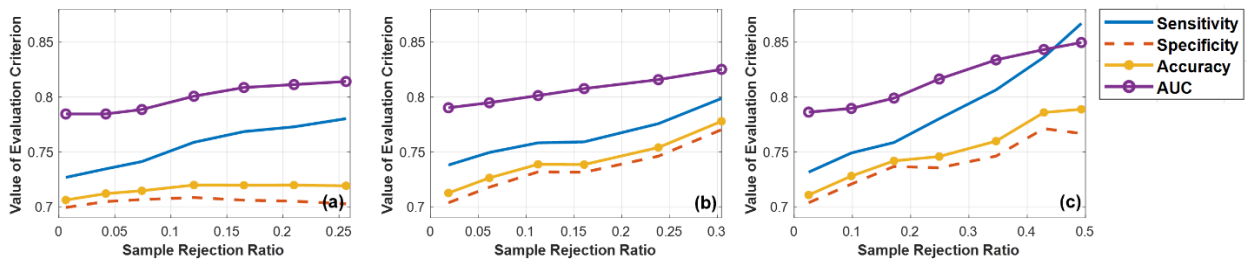

**Figure 4.** Delta-radiomics feature-based multi-classifier, multi-objective, and multi-modality (Delta-mCOM) model performance at different testing sample rejection ratios. (a) Reject prediction of testing samples based on epistemic uncertainty score; (b) Reject testing samples based on aleatoric uncertainty score; (c) Reject testing samples based on the combination of epistemic uncertainty and aleatoric uncertainty score.

**Table 2.** Post-treatment head and neck cancers locoregional recurrence prediction performance using delta-radiomics feature-based multi-classifier, multi-objective, and multi-modality (Delta-mCOM) model with different sample rejection methods and different rejection ratios. The P-values are obtained by comparing the ROCs of different methods to the combined uncertainty method at 25.0% rejection ratio and 49.3% rejection ratio.

| Rejection method | Rejection Ratio | Sensitivity | Specificity | Accuracy | AUC | P-value (vs. 25% comb.) | P-value (vs. 49% comb.) |
|---|---|---|---|---|---|---|---|
| None | 0% | 0.76±0.03 | 0.71±0.01 | 0.72±0.01 | 0.80±0.01 | 0.25 | <0.01 |
| Epistemic | 25.7% | 0.78±0.04 | 0.70±0.01 | 0.72±0.02 | 0.81±0.01 | 0.96 | 0.17 |
| Aleatoric | 23.9% | 0.77±0.03 | 0.74±0.02 | 0.75±0.02 | 0.81±0.01 | 0.87 | 0.12 |
| Combined | 25.0% | 0.78±0.04 | 0.73±0.02 | 0.75±0.02 | 0.82±0.01 | ---- | 0.35 |
| Combined | 49.3% | 0.87±0.03 | 0.77±0.02 | 0.79±0.02 | 0.85±0.02 | 0.35 | ---- |
| Rejected[b] | 49.3% | 0.56±0.14 | 0.57±0.05 | 0.57±0.04 | 0.63±0.09 | <0.01 | <0.01 |

[b] Rejected here refers to the rejected high-uncertainty patient group using the combined rejection method.

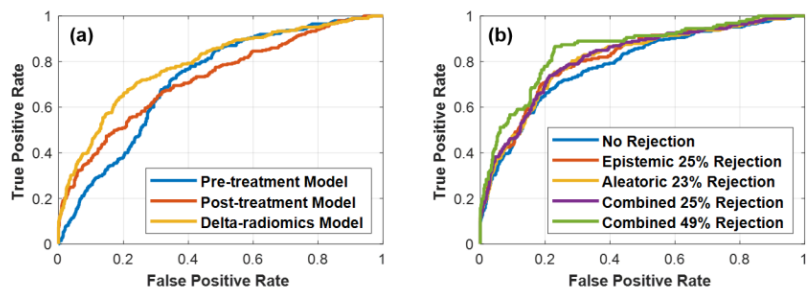

**Figure 5.** ROCs of head and neck cancers (HNC) locoregional recurrence (LRR) prediction models. (a) ROCs of the proposed delta-radiomics feature-based multi-classifier, multi-objective, and multi-modality (Delta-mCOM) models and mCOM models built with features collected at different time points; (b) Delta-mCOM models with different testing sample rejection methods for post-treatment HNC LRR prediction.

## 4. DISCUSSION

Early prediction/detection of LRR after radiotherapy may prompt early salvage therapy for HNC patients, while accurately identifying a complete response may prevent the need for unnecessary surgical intervention.[3,4] Therefore, a reliable post-treatment model for early HNC LRR prediction is of great importance to guide post-treatment patient management. This study employed BL-DRF features extracted from pre- and post-treatment PET/CT scans to make high-accuracy post-treatment LRR predictions. We applied then the LRO strategy to achieve more reliable predictions. Our software codes allowing to repeat the experiments and results presented in this work are freely shared under the GNU General Public License on the GitHub page: https://github.com/wangkaiwan/LRO_ReliabelRadiomics.

The use of BL-DRFs for post-treatment HNC LRR predictions was shown to help improve prediction accuracy. In our previous work, we proposed the mCOM model for pre-treatment HNC LRR predictions. Our model achieved AUC values of 0.77 and 0.75 on a publicly available and institutional dataset, respectively [8,10]. In this study, when we applied the post-treatment radiomics alone and retrained the model for post-treatment HNC LRR prediction, it achieved an AUC of 0.73. The ROC curves for the pre- and post-treatment predictions were not significantly different, and the performance of the post-treatment model was numerically lower than the pre-treatment model. Inspired by the clinical use of PET/CT for post-treatment patient management,[3-5] we calculated the difference between pre- and post-treatment radiomics features and concatenated them together with baseline radiomics feature as BL-DRF for post-treatment HNC LRR prediction. Our experiment showed the introduction of delta-radiomics significantly improved the performance (AUC=0.80) compared with models built with pre- or post-treatment radiomics features.

Models with multiple modalities yielded better AUC values than those built with single modalities in this study. These results suggest that the multiple modalities fusion strategy using ER could achieve more robust results. For example, the prediction power of the clinical feature sets used in this study was limited when it was used alone (AUC=0.67). Still, it would always provide improvement to the fused model performance.

In addition, we demonstrated the feasibility and efficiency of the LRO strategy for reliable HNC LRR prediction. In this work, our understanding of ''reliable outcome prediction'' is an outcome prediction method that is "aware" of its potential limitations. This model's "awareness" requires the predictor to estimate the reliability of each prediction in addition to giving the prediction result. And with both the reliability and prediction results, we hope to obtain better patient management decisions especially for challenging clinical cases. In our workflow, we used the proposed the Delta-mCOM model to make predictions. With each predicted, we calculated the aleatoric uncertainty via TTA, and epistemic uncertainty using anomaly detection models. Then we used the aleatoric and epistemic uncertainties in the combined rejection method as the estimation of reliability. The prediction results for the "high-reliability" patient group (around 50% rejection ratio) selected under the guidance of the LRO strategy achieved significantly higher performance (AUC=0.85) than both the rejected patient group (AUC=0.64) and the original testing group (AUC=0.80).

Although the Delta-mCOM model under the LRO strategy only achieves an AUC less than 0.90, we believe the reliable outcome prediction strategy with rejection option is an avenue for continued research. Two feasible ways to achieve more accurate predictions for treatment outcome prediction applications are upgrading the prediction model with new algorithms and collecting additional patient data. However, algorithm modifications alone might not be enough to improve a model's performance to be clinically acceptable. Moreover, given that data collection is expensive, time-consuming, and labor-intensive for most medical problems, this approach might not be practical. However, with the proposed reliable outcome prediction strategy, these issues can be eased. Take HNC LRR prediction as an example, through selecting an appropriate reliability threshold, we can find a group of patients that have very reliable predictions which may meet the clinical requirement. Physicians can feel more comfortable with the predictions for these select group of patients. For those where predictions are not made, then standard careful assessment by physicians, along additional medical tests, or more frequent surveillance, will be necessary. Again, this method will not improve the prediction performance on all the testing data, it makes the prediction procedure more clinically reliable.

Our study has several limitations. First, although similar imaging protocols were used for our patient cohort, the repeatability of radiomics features used was not analyzed, which might decrease the accuracy of delta-radiomics feature measurement and lead to reduced performance of the outcome prediction model.[16,29,42] A phantom experiment could be performed under the same imaging protocol used in this study. Features that remain the same or have a high Intra-Class Correlation Coefficient (ICC) can be selected for delta-radiomics feature calculation and prediction model construction. Second, to model the distribution of the radiomics features used for the delta-mCOM model training, we built serval anomaly detection networks with simple fully connected AE structures. We then used the anomaly scores from these anomaly detection networks as the estimation of epistemic uncertainties to measure the similarity between training and testing samples. To improve the prediction reliability of the selected low-uncertainty group, it is of importance for us to explore more suitable feature distribution modelling methods and feature similarity measurement methods for radiomics-based application. Third, we did not perform external validation with the proposed Delta-mCOM model and LRO strategy because there is no publicly available HNC patient dataset containing post-treatment images. When such a dataset is available, we should investigate the LRO utility. Forth, we empirically selected some of the hyperparameters, which includes size of pre-selected feature set, types of the used classifiers, and number of population and generation for the immune algorithm (mTO). The stability of the proposed model under different hyperparameter selection is a potential limitation of this study. Finally, we did not use deep learning methods to construct the prediction model. Although our current dataset is relatively small for constructing a deep learning model, transfer learning methods and attention-based learning methods reported as feasible to construct predictive models with a limited number of training samples might alleviate this limitation.[20,43,44] Deep learning-based outcome prediction using PET/CT images collected during HNC patient management is one of our future research directions, and how to apply the LRO strategy to it would be an interesting research topic.

## 5. CONCLUSIONS

Overall, we proposed a Delta-mCOM model for post-treatment HNC LRR prediction and introduced the LRO strategy to improve the reliability of prediction. The inclusion of the delta-radiomics features significantly improved the accuracy of HNC LRR prediction (AUC=0.80). Furthermore, the proposed Delta-radiomics model can give more reliable predictions on the selected testing patient group by rejecting high-uncertainty samples (AUC=0.85 with a rejection ratio of 50%). The proposed reliable outcome prediction method is worthy of further investigation on larger prospective cohorts and application on other treatment outcome prediction problems.

**ACKNOWLEDGMENTS**

We acknowledge funding support from National Institutes of Health (R01CA251792).

**CONFLICT OF INTEREST**

The authors have no relevant conflict of interest to disclose.

**SUPPORTING MATERIAL**

SupplementaryDocument.docx

**LIST OF FIGURES AND TABLES**

**Figure 1.** Illustration of unreliable predictions when a model is trained with limited data samples (modified from Ref. 22). When the query sample lies close to the decision boundary, the prediction is unreliable because of low prediction confidence. Additionally, the prediction can be unreliable despite a high class probability value when the query sample is far from the training data distribution.

**Figure 2.** Workflow of the proposed reliable delta-radiomics feature-based multi-classifier, multi-objective, and multi-modality (Delta-mCOM) model for post-treatment head and neck cancers (HNC) locoregional recurrence (LRR) prediction. (a) Workflow of Delta-mCOM model construction. Three separated single-

modality models (clinical model, PET-radiomics based model, and CT-radiomics based model) are trained individually and their out probabilities are fused together to generate the final prediction of Delta-mCOM model. (b) Workflow of the training and validation processes of a single-modality multi-objective multi-classifier prediction model. A trained model contains a set of different solutions $\{s_1, s_2, \ldots, s_N\}$ (Pareto-optimal solution set), each solution comprises of three trainable parameters: feature selection vector $f_i$, classifier parameter $\beta_i$, classifier weighting factor $w_i$, where $i = 1, 2, \ldots, I$ is the index of a solution, $I$ is the maximum size of the solution set, $N$ is the size of the Pareto-optimal solution set, and $N < I$. (c) Workflow of HNC LRR prediction using Delta-mCOM model with rejection option.

**Figure 3.** AUCs of **head and neck** cancers locoregional recurrence prediction with radiomics features collected from different modalities at different time points and the proposed delta-radiomics feature-based multi-classifier, multi-objective, and multi-modality (Delta-mCOM) model. Note that concatenation of baseline and delta-radiomics features (BL-DRFs) were used as the input for delta-radiomics-based models.

**Figure 4.** Delta-radiomics feature-based multi-classifier, multi-objective, and multi-modality (Delta-mCOM) model performance at different testing sample rejection ratios. (a) Reject prediction of testing samples based on epistemic uncertainty score; (b) Reject testing samples based on aleatoric uncertainty score; (c) Reject testing samples based on the combination of epistemic uncertainty and aleatoric uncertainty score.

**Figure 5.** ROCs of head and neck cancers (HNC) locoregional recurrence (LRR) prediction models. (a) ROCs of the proposed delta-radiomics feature-based multi-classifier, multi-objective, and multi-modality (Delta-mCOM) models and mCOM models built with features collected at different time points; (b) Delta-mCOM models with different testing sample rejection methods for post-treatment HNC LRR prediction.

**Table 1.** Results of head and neck cancers locoregional recurrence prediction with different features. P-values here are obtained by comparing the ROCs of different models to that of the proposed delta-radiomics feature-based multi-classifier, multi-objective, and multi-modality (Delta-mCOM) model (last row). Note

that concatenation of baseline and delta-radiomics features (BL-DRFs) were used as the input for delta-radiomics-based models.

**Table 2.** Post-treatment head and neck cancers locoregional recurrence prediction performance using delta-radiomics feature-based multi-classifier, multi-objective, and multi-modality (Delta-mCOM) model with different sample rejection methods and different rejection ratios. The P-values are obtained by comparing the ROCs of different methods to the combined uncertainty method at 25.0% rejection ratio and 49.3% rejection ratio.

## REFERENCES

1. Beswick DM, Gooding WE, Johnson JT, Branstetter BFt. Temporal patterns of head and neck squamous cell carcinoma recurrence with positron-emission tomography/computed tomography monitoring. *Laryngoscope.* 2012;122(7):1512-1517.
2. Oksuz DC, Prestwich RJ, Carey B, et al. Recurrence patterns of locally advanced head and neck squamous cell carcinoma after 3D conformal (chemo)-radiotherapy. *Radiat Oncol.* 2011;6:54.
3. Gupta T, Master Z, Kannan S, et al. Diagnostic performance of post-treatment FDG PET or FDG PET/CT imaging in head and neck cancer: a systematic review and meta-analysis. *European journal of nuclear medicine and molecular imaging.* 2011;38(11):2083-2095.
4. Goel R, Moore W, Sumer B, Khan S, Sher D, Subramaniam RM. Clinical practice in PET/CT for the management of head and neck squamous cell cancer. *American Journal of Roentgenology.* 2017;209(2):289-303.
5. Young H, Baum R, Cremerius U, et al. Measurement of clinical and subclinical tumour response using [18F]-fluorodeoxyglucose and positron emission tomography: review and 1999 EORTC recommendations. *European journal of cancer.* 1999;35(13):1773-1782.
6. Sheikhbahaei S, Taghipour M, Ahmad R, et al. Diagnostic accuracy of follow-up FDG PET or PET/CT in patients with head and neck cancer after definitive treatment: a systematic review and meta-analysis. *American Journal of Roentgenology.* 2015;205(3):629-639.
7. Kim J, Roh J, Kim J, et al. 18 F-FDG PET/CT surveillance at 3–6 and 12 months for detection of recurrence and second primary cancer in patients with head and neck squamous cell carcinoma. *British journal of cancer.* 2013;109(12):2973-2979.
8. Wang K, Zhou Z, Wang R, et al. A multi-objective radiomics model for the prediction of locoregional recurrence in head and neck squamous cell cancer. *Medical physics.* 2020;47(10):5392-5400.
9. Parmar C, Grossmann P, Rietveld D, Rietbergen MM, Lambin P, Aerts HJ. Radiomic machine-learning classifiers for prognostic biomarkers of head and neck cancer. *Frontiers in oncology.* 2015;5:272.
10. Vallieres M, Kay-Rivest E, Perrin LJ, et al. Radiomics strategies for risk assessment of tumour failure in head-and-neck cancer. *Sci Rep-Uk.* 2017;7.
11. Wong AJ, Kanwar A, Mohamed AS, Fuller CD. Radiomics in head and neck cancer: from exploration to application. *Translational Cancer Research.* 2016;5(4):371-382.

12. Zhou Z, Wang K, Folkert MR, et al. Multifaceted radiomics for distant metastasis prediction in head & neck cancer. *Physics in Medicine & Biology.* 2020.

13. Parmar C, Leijenaar RTH, Grossmann P, et al. Radiomic feature clusters and Prognostic Signatures specific for Lung and Head & Neck cancer. *Sci Rep-Uk.* 2015;5.

14. Lin P, Yang P-F, Chen S, et al. A Delta-radiomics model for preoperative evaluation of Neoadjuvant chemotherapy response in high-grade osteosarcoma. *Cancer Imaging.* 2020;20(1):1-12.

15. Alahmari SS, Cherezov D, Goldgof DB, Hall LO, Gillies RJ, Schabath MB. Delta radiomics improves pulmonary nodule malignancy prediction in lung cancer screening. *IEEE Access.* 2018;6:77796-77806.

16. Morgan HE, Wang K, Dohopolski M, et al. Exploratory ensemble interpretable model for predicting local failure in head and neck cancer: the additive benefit of CT and intra-treatment cone-beam computed tomography features. *Quantitative Imaging in Medicine and Surgery.* 2021.

17. Jeon SH, Song C, Chie EK, et al. Delta-radiomics signature predicts treatment outcomes after preoperative chemoradiotherapy and surgery in rectal cancer. *Radiation oncology.* 2019;14(1):1-10.

18. Shayesteh S, Nazari M, Salahshour A, et al. Treatment response prediction using MRI-based pre-, post-, and delta-radiomic features and machine learning algorithms in colorectal cancer. *Medical physics.* 2021;48(7):3691-3701.

19. Chen X, Zhou M, Wang Z, Lu S, Chang S, Zhou Z. Immunotherapy treatment outcome prediction in metastatic melanoma through an automated multi-objective delta-radiomics model. *Computers in Biology and Medicine.* 2021;138:104916.

20. Shen C, Nguyen D, Zhou Z, Jiang SB, Dong B, Jia X. An introduction to deep learning in medical physics: advantages, potential, and challenges. *Physics in Medicine & Biology.* 2020;65(5):05TR01.

21. Gillies RJ, Kinahan PE, Hricak H. Radiomics: Images Are More than Pictures, They Are Data. *Radiology.* 2016;278(2):563-577.

22. Gao J, Yao J, Shao Y. Towards Reliable Learning for High Stakes Applications. Paper presented at: Proceedings of the AAAI Conference on Artificial Intelligence2019.

23. Chen LY, Zhou ZG, Sher D, et al. Combining many-objective radiomics and 3D convolutional neural network through evidential reasoning to predict lymph node metastasis in head and neck cancer. *Phys Med Biol.* 2019;64(7).

24. Zhou ZG, Folkert M, Iyengar P, et al. Multi-objective radiomics model for predicting distant failure in lung SBRT. *Phys Med Biol.* 2017;62(11):4460-4478.

25. Chiesa S, Bartoli FB, Longo S, et al. Delta Radiomics Features Analysis for the Prediction of Patients Outcomes in Glioblastoma Multiforme: The Generating Hypothesis Phase of GLIFA Project. *Int J Radiat Oncol.* 2018;102(3):S213-S213.

26. Fave X, Zhang LF, Yang JZ, et al. Delta-radiomics features for the prediction of patient outcomes in non-small cell lung cancer. *Sci Rep-Uk.* 2017;7.

27. Jeon SH, Song C, Chie EK, et al. Delta-radiomics signature predicts treatment outcomes after preoperative chemoradiotherapy and surgery in rectal cancer. *Radiat Oncol.* 2019;14(1):43.

28. Cortes C, DeSalvo G, Mohri M. Learning with rejection. Paper presented at: International Conference on Algorithmic Learning Theory2016.

29. Zwanenburg A, Vallières M, Abdalah MA, et al. The image biomarker standardization initiative: standardized quantitative radiomics for high-throughput image-based phenotyping. *Radiology.* 2020;295(2):328-338.

30. Peng HC, Long FH, Ding C. Feature selection based on mutual information: Criteria of max-dependency, max-relevance, and min-redundancy. *Ieee T Pattern Anal.* 2005;27(8):1226-1238.

31. Chawla NV, Bowyer KW, Hall LO, Kegelmeyer WP. SMOTE: synthetic minority over-sampling technique. *Journal of artificial intelligence research.* 2002;16:321-357.

32. Bartlett PL, Wegkamp MH. Classification with a Reject Option using a Hinge Loss. *Journal of Machine Learning Research.* 2008;9(8).

33. Senge R, Bösner S, Dembczyński K, et al. Reliable classification: Learning classifiers that distinguish aleatoric and epistemic uncertainty. *Information Sciences.* 2014;255:16-29.

34. Jakeman J, Eldred M, Xiu D. Numerical approach for quantification of epistemic uncertainty. *Journal of Computational Physics.* 2010;229(12):4648-4663.

35. Kendall A, Gal Y. What uncertainties do we need in bayesian deep learning for computer vision? *arXiv preprint arXiv:170304977.* 2017.

36. Zhao Y, Nasrullah Z, Li Z. Pyod: A python toolbox for scalable outlier detection. *arXiv preprint arXiv:190101588.* 2019.

37. Baur C, Wiestler B, Albarqouni S, Navab N. Deep autoencoding models for unsupervised anomaly segmentation in brain MR images. Paper presented at: International MICCAI Brainlesion Workshop2018.

38. Ayhan MS, Berens P. Test-time data augmentation for estimation of heteroscedastic aleatoric uncertainty in deep neural networks. 2018.

39. Dohopolski M, Chen L, Sher D, Wang J. Predicting lymph node metastasis in patients with oropharyngeal cancer by using a convolutional neural network with associated epistemic and aleatoric uncertainty. *Physics in Medicine & Biology.* 2020;65(22):225002.

40. Sun X, Xu W. Fast implementation of DeLong's algorithm for comparing the areas under correlated receiver operating characteristic curves. *IEEE Signal Processing Letters.* 2014;21(11):1389-1393.

41. Team RC. R: A language and environment for statistical computing. 2013.

42. Zwanenburg A. Radiomics in nuclear medicine: robustness, reproducibility, standardization, and how to avoid data analysis traps and replication crisis. *European journal of nuclear medicine and molecular imaging.* 2019;46(13):2638-2655.

43. Chen L, Dohopolski M, Zhou Z, et al. Attention Guided Lymph Node Malignancy Prediction in Head and Neck Cancer. *International Journal of Radiation Oncology* Biology* Physics.* 2021;110(4):1171-1179.

44. Chen S, Ma K, Zheng Y. Med3d: Transfer learning for 3d medical image analysis. *arXiv preprint arXiv:190400625.* 2019.